\documentclass[twocolumn,showpacs,superscriptaddress,preprintnumbers,amsmath,amssymb]{revtex4}

\usepackage{graphicx}
\usepackage{dcolumn}
\usepackage{bm}

\makeatletter 

\newcommand{\Rmnum}[1]{\expandafter\@slowromancap\romannumeral #1@}
\makeatother

\begin{document}


\title{Solving the inverse problem of noise-driven dynamic networks}

\author{Zhaoyang Zhang}
\affiliation{Department of Physics, Beijing Normal University, Beijing 100875, China}
\author{Zhigang Zheng}
\affiliation{Department of Physics, Beijing Normal University, Beijing 100875, China}
\author{Haijing Niu}
\affiliation{State Key Laboratory of Cognitive Neuroscience and Learning and International Digital Group (IDG)/McGovern Institute for Brain Research, Beijing Normal University, Beijing 100875, China}
\affiliation{Center for Collaboration and Innovation in Brain and Learning Sciences, Beijing Normal University, Beijing 100875, China}
\author{Yuanyuan Mi}
\affiliation{State Key Laboratory of Cognitive Neuroscience and Learning and International Digital Group (IDG)/McGovern Institute for Brain Research, Beijing Normal University, Beijing 100875, China}
\author{Si Wu}
\affiliation{State Key Laboratory of Cognitive Neuroscience and Learning and International Digital Group (IDG)/McGovern Institute for Brain Research, Beijing Normal University, Beijing 100875, China}
\affiliation{Center for Collaboration and Innovation in Brain and Learning Sciences, Beijing Normal University, Beijing 100875, China}
\author{Gang Hu}
\email{ganghu@bnu.edu.cn}
\affiliation{Department of Physics, Beijing Normal University, Beijing 100875, China}

\date{\today}

\begin{abstract}
Nowadays massive amount of data are available for analysis in natural and social systems. Inferring system structures from the data, i.e., the inverse problem, has become one of the central issues in many disciplines and interdisciplinary studies. In this Letter, we study the inverse problem of stochastic dynamic complex networks. We derive analytically a simple and universal inference formula called double correlation matrix (DCM) method. Numerical simulations confirm that the DCM method can accurately depict both network structures and noise correlations by using available kinetic data only. This inference performance was never regarded possible by theoretical derivation, numerical computation and experimental design.
\end{abstract}

\maketitle

\emph{Introduction.} In recent decades, large scale of data sets have been accumulated in various and wide fields, in particular in social and biological systems
~\cite{Butte2000,Kim2001,Bar-Joseph2003,Bar-Joseph2012}. There are massive amount of data available for utilization, however, the system structures yielding these data are often not clear~\cite{Feist2009,DeSmet2010}. Therefore, deducing the connectivity of systems from these data, i.e., the inverse problem, turns to be today one of the central issues in interdisciplinary fields~\cite{Yeung2002,Stuart2003,Segal2003,Hu2007,Lezon2006,Barzel2013,Feizi2013}. A typical example of inference efforts is a recent project of the Dialogue on Reverse Engineering Assessment and Methods (DREAM) which has attracted extensive attention for reconstructing gene regulatory networks from high-throughput microarray data~\cite{Marbach2010,Marbach2012}. Similar goals have been also pursued in other fields, such as neural networks~\cite{Bullmore2009}, ecosystems~\cite{Sugihara2012}, chemical reactions \cite{Arkin1995,Arkin1997} and so on. Most of biological and social systems contain many units which evolve collectively with very complicated interaction structures represented by complex networks~\cite{Watts1998,Barabasi1999,Barabasi2004}. Mathematically, the dynamics of these complex systems are extensively described by sets of coupled ordinary differential equations (ODEs)~\cite{Glass,Goldbeter,alon,Tsai}. The inverse problems of these systems can thus be interpreted as to retrieve the interaction Jacobian matrices from the measurable data of dynamical variables of networks. So far, a wide range of network inference methods have been proposed to address this issue in diverse fields. Available methods can be classified into several broad categories~\cite{Basso2005,Bansal2007,Marbach2012,Villaverde2013}: Bayesian networks and probabilistic graphical models, which maximize a scoring function over alternative network models~\cite{Jansen2003,Friedman2004}; regression techniques, which fit the data to a priori models~~\cite{Haury}; integrative bioinformatics approaches, which combine data from a number of independent experimental clues~\cite{Gardner2003,Covert2004}; statistical
methods, which rely on a variety of measures of pairwise correlations or mutual information and other methods~\cite{Eisen1998,Basso2005,Faith2007}.

The complexity of networks can hinder the attempt to solve the inverse problems~\cite{Gardner2003}. Moreover, the network dynamics are inevitably perturbed by many uncontrollable impacts, called noise, and these random and unknown perturbations make the inverse problems even more difficult. Actually, noise can play two seemingly contradictory effects. On one hand noise can contaminate data, mask noise-free network dynamics and thus lead to inference errors. On the other hand, noise perturbations are helpful to provide rich distinctive data which involve useful information for effective inferences, as emphasized recently by~\cite{Ren2010}. However, the latter role of noise has been ignored by most of current inference methods. The results of the currently prevailing inference methods are thus unsatisfactory, in particular if noise is unknown and noise effect plays crucial role in data productions. To overcome these difficulties, new comprehensive physical ideas and intelligent mathematical methods become absolutely necessary.

In the present work, a novel double correlation matrix (DCM) method is proposed to generally solve the inverse problems of dynamic complex networks driven by noise, and we derive a compact and universal algorithm $\hat{\mathbf{A}}= \hat{\mathbf{B}}\hat{\mathbf{C}}^{-1}$ with $\hat{\mathbf{A}}$ being the target of the inverse problems, i.e., the interaction Jacobian matrix, $\hat{\mathbf{C}}$ the variable-variable correlation matrix and $\hat{\mathbf{B}}$ the velocity-variable correlation matrix. All elements in $\hat{\mathbf{B}}$ and $\hat{\mathbf{C}}$ can be explicitly computed from the measurable variable data only.

In contrast with all the previous methods, the DCM method has two remarkable advantages. First, it extracts more useful information from the available data by computing double correlation matrices $\hat{\mathbf{B}}$ and $\hat{\mathbf{C}}$ while only a single matrix has been considered in most of inference methods~\cite{Levnaji2014}. Second, it effectively filters out noise contamination without requiring detailed correlation statistics of noise by using the fast varying property of noise. This property is available for most of practical systems while, to our knowledge, has never been fully utilized so far in inverse computations. Due to these advantages, the DCM method can infer the structure of noise-driven dynamic networks incomparably more effectively and accurately than currently prevailing inference methods do.

\emph{Theory.} A large class of dynamic networks can be most generally represented by the following coupled ODEs driven by noise
\begin{equation}
\dot{\mathbf{x}}(t)=\mathbf{f}(\mathbf{x}(t),\eta(t)),
\label{eq:02}
\end{equation}
with variables ${\mathbf{x}(t)=(x_{1}(t),x_{2}(t),\cdots,x_{N}(t))}$, noise ${\mathbf{\eta}(t)=(\eta_{1}(t),\eta_{2}(t),\cdots,\eta_{M}(t))}$ and dynamic fields ${\mathbf{f}(\mathbf{x}(t),\eta(t))=(f_{1}(\mathbf{x}(t),\eta(t)),\cdots,f_{N}(\mathbf{x}(t),\eta(t)))}$. In this Letter, we adopt white noise approximation for very short noise correlation time,
$$\langle{\eta_{i}(t)}\rangle=0, {\langle{\eta_{i}(t)\eta_{j}(t')}\rangle}=D_{ij}\delta(t-t'), {i, j=1,2,\cdots,M}.$$
In most of realistic systems, this approximation does be valid when noise serves as perturbations from microscopic world varying much faster than the macroscopic variables. Around any phase space point $\mathbf{x}$ and for small noise approximation, Eq.~(\ref{eq:02}) can be linearized to
\begin{equation}
\dot{\mathbf{y}}(t)=\hat{\mathbf{A}}(\mathbf{x})\mathbf{y}(t)+\Gamma(t),
\label{eq:04}
\end{equation}
$${{\mathbf{y}}(t)=(y_{1}(t),\cdots,y_{N}(t))}, {\Gamma(t)=(\Gamma_{1}(t),\cdots,\Gamma_{N}(t))},$$
\begin{equation}
{\langle{\Gamma(t)}\rangle=0}, {\langle{\Gamma_{i}(t)\Gamma_{j}(t')}\rangle=Q_{ij}\delta(t-t')}.
\label{eq:05}
\end{equation}
where ${\mathbf{y}(t)=\mathbf{x}(t)-\mathbf{x}},$ ${A_{ij}(\mathbf{x})=\frac{\partial{f_{i}}}{\partial{x_j}}\mid_{\mathbf{x}(t)=\mathbf{x},\eta(t)=0}},$ $\Gamma_{i}(t)={\sum_{\mu=1}^{M}}G_{i{\mu}}{\eta_{\mu}},$ ${G_{i\mu}=\frac{\partial{f_{i}}}{\partial{\eta_{\mu}}}\mid_{\mathbf{x}(t)=\mathbf{x},\eta(t)=0}}$ and ${Q_{ij}={\sum_{\mu=1}^M}{\sum_{\nu=1}^M}G_{i\mu}D_{\mu{\nu}}G_{\nu{j}}^{T}}.$

Without noise ($\eta(t)=0$) Eq.~(\ref{eq:02}) evolves to one of their attractors over time which may be a stable steady state, a periodic or chaotic state. In any case, the dimension of the attractor $d_{a}$ should be considerably smaller than that of the original network $d_{a}\ll{N}$. Therefore, it is impossible to use the data set of an attractor of noise-free system to infer the network structure due to lack of sufficient information in the data. Existence of noise can scatter the variable data to fill $N$-dimensional phase space and provide possibility (sufficient information) to identify the full interactions of the network.

\begin{figure}[h!]
\resizebox{8.0cm}{6.0cm}{\includegraphics{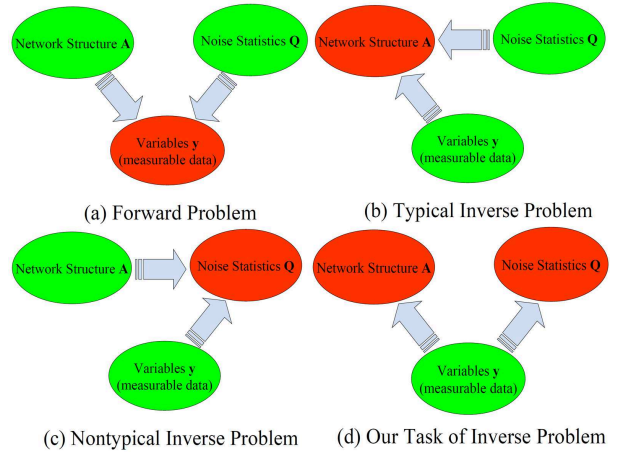}}
\caption{\label{Fig1} (Color online) Schematic figures on forward and inverse problems of complex networks of Eq.~(\ref{eq:04}). (a) Forward problem of Eq.~(\ref{eq:04}): From network structure $\hat{\mathbf{A}}$ and noise statistics $\hat{\mathbf{Q}}$ to calculate output variable data $\mathbf{y}(t)$. (b) Typical inverse problem: From available output variable data $\mathbf{y}(t)$ and noise statistics $\hat{\mathbf{Q}}$ to retrieve network Jacobian structure $\hat{\mathbf{A}}$, by using matrix equation $\hat{\mathbf{A}}\hat{\mathbf{C}}+\hat{\mathbf{C}}\hat{\mathbf{A}}^{T}=-\hat{\mathbf{Q}}$~\cite{Ren2010,Ching2013,Risken} with $\hat{\mathbf{A}}$, $\hat{\mathbf{Q}}$ and $\hat{\mathbf{C}}$ given in Eqs.~(\ref{eq:04})(\ref{eq:05})(\ref{eq:11}). (c) Another nontypical inverse problem: From available variable data $\mathbf{y}(t)$ and Jacobian structure $\hat{\mathbf{A}}$ to depict noise statistics $\hat{\mathbf{Q}}$. (d) A desirable inverse computational target not regarded possible so far: From available variable data $\mathbf{y}(t)$ only to reveal both network Jacobian structure $\hat{\mathbf{A}}$ and noise statistics $\hat{\mathbf{Q}}$. And this is the task of this Letter.}
\end{figure}

With both linearized matrix $\hat{\mathbf{A}}$ in (\ref{eq:04}) and noise statistics $\hat{\mathbf{Q}}$ in (\ref{eq:05}) given, we can calculate output variables $\mathbf{y}(t)$ as a well known forward problem of dynamic networks (Fig.~\ref{Fig1}(a)). A typical inverse problem is to retrieve the interaction Jacobian matrix $\hat{\mathbf{A}}$ with measurable output data $\mathbf{y}(t)$ and known noise statistics $\hat{\mathbf{Q}}$ (Fig.~\ref{Fig1}(b))~\cite{Ren2010,Risken}. This inverse problem can be solved for certain simple symmetric network structures, whereas it is not solvable in many complicated cases, such as when networks have asymmetric links or when the basic noise-free networks have nonsteady and nonsynchronous motions~\cite{Ching2013,Wang2009,Ren2010}. Another trivial and nontypical inverse problem is to reveal noise statistics $\hat{\mathbf{Q}}$ with known output data $\mathbf{y}(t)$ and network structures $\hat{\mathbf{A}}$ (Fig.~\ref{Fig1}(c)). However, the inference condition of Fig.~\ref{Fig1}(b) is not reasonable in practice. Since noise represents some random and uncontrollable factors, even less information can be obtained on noise than on network structures and it is unreasonable to have known knowledge of noise statistics $\hat{\mathbf{Q}}$ to infer unknown network structure $\hat{\mathbf{A}}$. A most effective as well as most desirable inverse target is presented in Fig.~\ref{Fig1}(d) where one can solve the inverse problem merely from the measurable data $\mathbf{y}(t)$ with both network structures $\hat{\mathbf{A}}$ and noise statistics $\hat{\mathbf{Q}}$ unknown. This inference performance has never been regarded possible so far by theoretical analysis and experimental design. And this is right the issue discussed in this Letter.

Now, we consider how to infer network structure $\hat{\mathbf{A}}$ merely from measurable data $\mathbf{y}(t)$. Suppose we have $L$ pairs of variable data ($\mathbf{x}(t_{q}), \mathbf{x}(t_{q}+\Delta{t_{q}}), {q=1,2,\cdots,L}$) in a small phase space region, with all $\Delta{t_{q}}$, $\|\mathbf{x}(t_{q'})-\mathbf{x}(t_{q})\|$, $\|\mathbf{x}(t_{q'}+\Delta{t_{q'}})-\mathbf{x}(t_{q})\|$ $\ll{1}$ for all $q$, $q'$. From the available data we can extract full information of $\mathbf{y}(t_{q})=\mathbf{x}(t_{q})-\mathbf{x}$ and $\mathbf{x}=\frac{1}{L}\sum_{q=1}^{L}{\mathbf{x}(t_{q})}$. The velocity term $\dot{\mathbf{y}}(t_{q})$ in Eq.~(\ref{eq:04}) can be measured as
\begin{equation}
\dot{y}_{i}(t_{q})=\frac{y_{i}(t_{q}+\Delta{t_{q}})-y_{i}(t_{q})}{\Delta{t_{q}}}.
\label{eq:07}
\end{equation}
With all the quantities ${\mathbf{y}}(t_{q})$ and $\dot{\mathbf{y}}(t_{q})$ measured, we can derive some explicit and compact algorithms from Eq.~(\ref{eq:04}) as
\begin{equation}
\hat{\mathbf{A}}=\hat{\mathbf{B}}\hat{\mathbf{C}}^{-1},
\label{eq:08}
\end{equation}
and
\begin{equation}
\hat{\mathbf{Q}}=-2\hat{\mathbf{B}}_{s}=-(\hat{\mathbf{B}}+\hat{\mathbf{B}}^{T}),
\label{eq:10}
\end{equation}
where $\hat{\mathbf{C}}=\langle{{\mathbf{y}}\mathbf{y}^{T}}\rangle$ and $\hat{\mathbf{B}}=\langle{\dot{\mathbf{y}}\mathbf{y}^{T}}\rangle$ are the variable-variable and velocity-variable correlation matrices, respectively,
\begin{equation}
C_{ij}={\frac{1}{L}}{\sum_{q=1}^{L}}y_{i}(t_{q})y_{j}(t_{q}), B_{ij}={\frac{1}{L}}{\sum_{q=1}^{L}}\dot{y}_{i}(t_{q})y_{j}(t_{q}).
\label{eq:11}
\end{equation}
And $\hat{\mathbf{B}}_{s}$ and $\hat{\mathbf{B}}^{T}$ are the symmetric part and the transposition of $\hat{\mathbf{B}}$, respectively. The detailed derivations of (\ref{eq:08}) and (\ref{eq:10}) are given in Supplemental Material SM~\uppercase\expandafter{\romannumeral1}.

Now, a novel double correlation matrix (DCM) method is proposed to generally solve the inverse problem of Eq.~(\ref{eq:04}) and also explicitly depict noise statistical correlation matrix $\hat{\mathbf{Q}}$, by using the simple and unified algorithms (\ref{eq:08}) and (\ref{eq:10}). All the targets of Fig.~\ref{Fig1}(d) are satisfactorily reached. Three points about formula (\ref{eq:08}) should be emphasized. First, the entire computation of (\ref{eq:08}) is merely based on the measurable output variable data $\mathbf{y}(t)$, and no additional information on network structure $\hat{\mathbf{A}}$ and noise correlations $\hat{\mathbf{Q}}$ are required. Second, correlation matrix $\hat{\mathbf{C}}$ has been extensively used by various inference methods, while correlation matrix $\hat{\mathbf{B}}$ has been rarely considered. In particular, no method has jointly used these double matrices in inference computations. Taking this advantage, Eq.~(\ref{eq:08}) can extract more information (information of both variable and velocity of variable) from available data than all else inference methods do, and this is the reason why we can achieve seemly impossible goals. Third, algorithms (\ref{eq:07}) and (\ref{eq:11}) for computing matrix $\hat{\mathbf{B}}$ are crucial for the DCM method to effectively filter out noise and deduce both $\hat{\mathbf{A}}$ and $\hat{\mathbf{Q}}$ without knowing any knowledge on noise term $\hat{\mathbf{Q}}$ (see SM~\uppercase\expandafter{\romannumeral1}).

\begin{figure}[h!]
\resizebox{8.0cm}{9.0cm}{\includegraphics{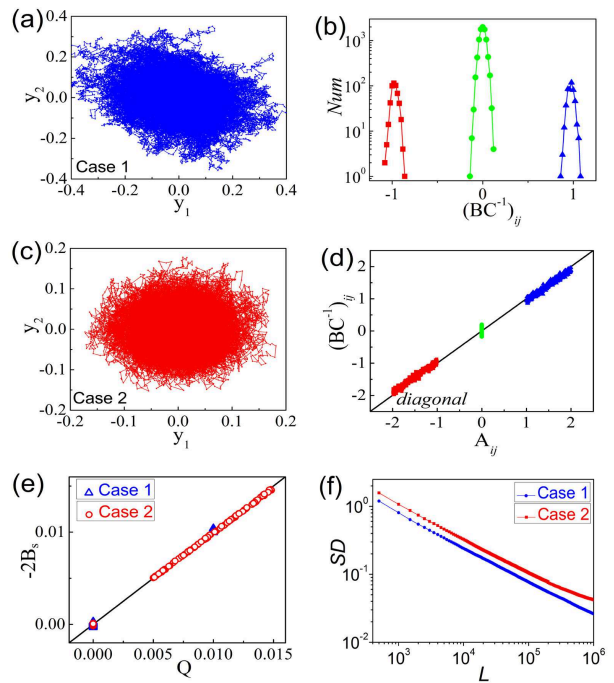}}
\caption{\label{Fig2} (Color online) Solving the inverse problem by double correlation matrix (DCM) method of algorithms~(\ref{eq:08}) and (\ref{eq:10}). $N=100$. $500$ active links and $500$ repressive ones are randomly chosen, and $A_{ij}(0)=0$ for all other off-diagonal matrix elements. $L=5\times{10^{5}}$ samples are yielded for inferences. (a)(b) Case 1: Jacobian matrix $\hat{\mathbf{A}}$ is given as: active interactions $A(a)_{ij}=1$; repressive interactions $A(r)_{ij}=-1$; the diagonal terms are set to $A_{ii}=-3$; and noise correlation matrix is given as $\hat{\mathbf{Q}}_{ij}=\sigma_{i}\delta_{ij}, \sigma_{i}=0.01$. (c)(d) Case 2: The same as (a)(b) with heterogeneous $\hat{\mathbf{A}}$ and $\hat{\mathbf{Q}}$, i.e., $A_{ij}(a)\in(1.0,2.0)$ and $A_{ij}(r)\in(-2.0,-1.0)$ with uniform distributions, $A_{ii}=-5$, and $\sigma_{i}\in(0.005,0.015)$ with uniform distribution. (a)(c) The trajectories in a $2D$ phase plane for the two sets of networks and parameters. (b) The distribution of off-diagonal elements of $\hat{\mathbf{A}}=\hat{\mathbf{B}}\hat{\mathbf{C}}^{-1}$ (blue triangles for $A_{ij}(a)$, red squares for $A_{ij}(r)$, and green circles for $A_{ij}(0)$) calculated by Eq.~(\ref{eq:08}) from data of (a). The computational elements of matrix $\hat{\mathbf{B}}\hat{\mathbf{C}}^{-1}$ coincide very well with the actual $\hat{\mathbf{A}}$, and all active, repressive and null interactions can be accurately predicted. (d) Matrix elements computed by $\hat{\mathbf{B}}\hat{\mathbf{C}}^{-1}$ are plotted against true Jacobian matrix elements $A_{ij}$ for Case 2 (all colors have the same meaning as (b)). All dots locate closely near the diagonal line, indicating correct inferences of both interaction natures and coupling strengths. (e) Element values of $-2\hat{\mathbf{B}}_{s}$ plotted versus element values of actual $\hat{\mathbf{Q}}$ for both the noises of (a)(c). The agreements between $-2\hat{\mathbf{B}}_{s}$ and $\hat{\mathbf{Q}}$ convincingly show that the DCM method can depict not only Jacobian matrix $\hat{\mathbf{A}}$ but also noise correlation matrix $\hat{\mathbf{Q}}$ merely from the output variable data. The task addressed in Fig.~\ref{Fig1}(d) is fulfilled indeed. (f) Standard deviation ($SD$) of Eq.~(\ref{eq:14}) plotted against the number of samples $L$. $SD$s decay as $\frac{1}{\sqrt{L}}$. From this law we can predict the necessary sample numbers for different inference precisions. }
\end{figure}

\emph{Computational results.} Equation~(\ref{eq:04}) can be generally derived for any phase space point where output data are available and $\hat{\mathbf{A}}$ is thus $\mathbf{x}$ dependent. Around a stable steady state of noise-free system, we can linearize Eq.~(\ref{eq:02}) directly around the fixed point ${\mathbf{x}}_{0}$ and set ${\mathbf{x}}={\mathbf{x}}_{0}$ and $\mathbf{y}(t)=\mathbf{x}(t)-{\mathbf{x}}_{0}$ for computing (\ref{eq:07})-(\ref{eq:11}). Now, our task is to reveal both Jacobian matrix $\hat{\mathbf{A}}$ in (\ref{eq:04}) and noise correlation $\hat{\mathbf{Q}}$ in (\ref{eq:05}) from measurable variable data $\mathbf{y}(t)$.

In Fig.~\ref{Fig2} we consider two examples for numerical simulations. The network size in Figs.~\ref{Fig2}(a)(b) (Case 1) is $N=100$. Among $10^{4}$ links there are $500$ active links $A_{ij}(a)=1$ and $500$ repressive ones $A_{ij}(r)=-1$ arbitrarily chosen, $A_{ij}(0)=0$ for all else off-diagonal matrix elements, and the diagonal terms are set to $A_{ii}=-3$ for keeping the network evolution bounded. Moreover, we take $\hat{\mathbf{Q}}_{ij}=\sigma_{i}\delta_{ij}, \sigma_{i}=0.01$. Running system (\ref{eq:04}) from $y_{i}(t=0)=0$ we produce $L=5\times10^{5}$ sets of data. Assume we know nothing about network structure $\hat{\mathbf{A}}$ and noise statistics $\hat{\mathbf{Q}}$ but only the data sequences of $\mathbf{y}(t)$, among which a projective trajectory is plotted in a $2D$ ($y_{1}(t)$, $y_{2}(t)$) phase plane in Fig.~\ref{Fig2}(a), which seems fully disordered. All the elements of matrices $\hat{\mathbf{B}}$ and $\hat{\mathbf{C}}$ can be computed from the data $\mathbf{y}(t)$ with Eqs.~(\ref{eq:07})(\ref{eq:11}), and then matrix $\hat{\mathbf{A}}$ can be retrieved by Eq.~(\ref{eq:08}). The results are presented in Fig.~\ref{Fig2}(b) where interactions depicted by $\hat{\mathbf{B}}\hat{\mathbf{C}}^{-1}$ agree well with the actual interactions $\hat{\mathbf{A}}$. In Figs.~\ref{Fig2}(c)(d) (Case 2), we do the same as Figs.~\ref{Fig2}(a)(b) with the noise statistics changed to $\hat{\mathbf{Q}}_{ij}=\sigma_{i}\delta_{ij}$ with $\sigma_{i}\in(0.005,0.015)$, and $A_{ij}$ set to $A_{ij}(a)\in(1.0,2.0)$, $A_{ij}(r)\in(-2.0,-1.0)$ (all randomly chosen with uniform distributions in their ranges) and the diagonal terms are set to $A_{ii}=-5$. Though the trajectory behaviors of Figs.~\ref{Fig2}(a) and \ref{Fig2}(c) deviate from each other substantially due to different noise correlations $\hat{\mathbf{Q}}$ and Jacobian matrix $\hat{\mathbf{A}}$, and these different data sets can surely yield considerably different matrices $\hat{\mathbf{B}}$ and $\hat{\mathbf{C}}$, it is remarkable that in Figs.~\ref{Fig2}(b) and \ref{Fig2}(d) the DCM method can correctly deduce both interaction matrices $\hat{\mathbf{A}}$ by applying the same algorithm $\hat{\mathbf{A}}=\hat{\mathbf{B}}\hat{\mathbf{C}}^{-1}$. On the contrary, most of inference methods use only correlations of $\hat{\mathbf{C}}$ (or other related quantities) and the results of these methods are thus seriously influenced by noise, and can never produce correct inferences with a universal formula for different noise correlations.

For confirming the conclusions of Eq.~(\ref{eq:10}), we plot $-2\hat{\mathbf{B}}_{s}$ computed from the variable data against actual noise statistics $\hat{\mathbf{Q}}$ in Fig.~\ref{Fig2}(e) for the two data sets of Figs.~\ref{Fig2}(a) and \ref{Fig2}(c). All these dots locate very closely around the diagonal line, convincingly justifying the prediction of Eq.~(\ref{eq:10}). The results of Eqs.~(\ref{eq:08})(\ref{eq:10}) are exact in the limits of white noise and $L\rightarrow{\infty}$, $\Delta{t_{q}}\rightarrow0$. In Fig.~\ref{Fig2}(f) we use the systems of Figs.~\ref{Fig2}(a) and \ref{Fig2}(c) to numerically compute the standard deviation ($SD$) of inferred values of interactions defined as
\begin{equation}
SD(L)=\sqrt{\frac{\sum_{i=1}^{N}\sum_{j=1}^{N}{([BC^{-1}]_{ij}-A_{ij})^{2}}}{N^2}},
\label{eq:14}
\end{equation}
where summations of $i$ and $j$ run over all matrix elements and $L$ is the total number of computational samples. It is clear that $SD{\propto}{\frac{1}{\sqrt{L}}}$, agreeing with the conclusion of exact inference solutions for $L\rightarrow{\infty}$. The theoretical conclusion that algorithms (\ref{eq:08}) and (\ref{eq:10}) can reveal both interaction structure $\hat{\mathbf{A}}$ and noise statistics $\hat{\mathbf{Q}}$ are remarkably verified by numerical simulations based on which we can, for the first time, reconstruct the stochastic dynamic networks Eq.~(\ref{eq:04}) from their variable outputs only. And this capacity of the DCM method is unique in all known inference methods.

In Fig.~\ref{Fig2} we study the inverse problem of noise-driven randomly constructed networks around stable fixed points. The DCM method can be generally applicable to various dynamic networks described by coupled stochastic ODEs, i.e., to different noise-free states such as periodic oscillations (SM~\uppercase\expandafter{\romannumeral2}) and even chaotic states (SM~\uppercase\expandafter{\romannumeral3}); to different network topologies such as scale-free networks (SM~\uppercase\expandafter{\romannumeral3}).

The currently prevailing inference methods are based on different information included in the output data to infer network structures. In SM~\uppercase\expandafter{\romannumeral4}, three commonly-used methods (Pearson correlation, Mutual information and Regression) are introduced for comparisons with the DCM method. For the noise-generated data, the results of the DCM method are considerably better than those of all the three commonly-used methods in both qualitative and quantitative predictions.

\emph{Conclusion.} In conclusion, we proposed a double correlation matrix (DCM) method to infer noise-driven dynamic networks from their output data. For given output time sequences yielded by stochastic network dynamics, the DCM method can accurately depict network structures with a compact formula. The method can depict not only the qualitative features of network structures (e.g., active, repressive and null natures of interactions), but also the precise strengths of interactions; not only the interaction Jacobian matrix $\hat{\mathbf{A}}$, but also the noise correlations $\hat{\mathbf{Q}}$. These are far beyond the capacity of all known inference methods.

There are two major ingredients enable the advantages of the DCM method. First, this method can extract more information because it measures not only the available data but also the variation velocities of variables while in most of inference methods only the former data have been used. The joint application of these two types of data makes the DCM method capable to infer network interactions much more accurately than other existing methods (see SM~\uppercase\expandafter{\romannumeral4}). Moreover, the DCM method uses the fast varying property of white noise (which is valid in most of realistic systems) so that matrix $\hat{\mathbf{B}}$ in Eq.~(\ref{eq:08}) can filter out noise effectively and then infer network structures without any knowledge of noise statistics (Fig.~\ref{Fig1}(d)). This has never been regarded possible so far.

Some conditions are required for the DCM method. The data should contain information of velocities of variables for computing matrix $\hat{\mathbf{B}}$. For doing so sufficiently fast data measurements are required. Since noise plays crucial role in yielding data, sufficiently large data sets are necessary for filtering out the noise contaminations. Many practically important systems can fulfill these conditions, among which brain networks and financial networks (e.g., stock market evolutions) are the most interesting candidates. In both cases, noises are often crucial in generating activity data and various quickly developed techniques guarantee high-frequency and non-invasive measurements and huge data collections. It is our further works to analyze these data sets to depict the possibly hidden network structures from dynamic variable data by applying the DCM method.

\emph{Acknowledgments.} This work is supported by National Foundation of Natural Science of China Grants 11135001 and 11174034 (to G.H.), 11075016 (to Z.Z.), 11305112 (to Y.M.), and 91132702 and 31261160495 (to S.W.); the Open Research Fund of the State Key Laboratory of Cognitive Neuroscience and Learning (CNLYB1211); and Natural Science Foundation of Jiangsu Province BK20130282.

\bibliography{apssamp}

\end{document}



\title{Solving the inverse problem of noise-driven dynamic networks}

\maketitle

\section{Analytical derivation of Eqs.~(5) and (6)}

Multiplying $\mathbf{y}^{T}(t)$ to both sides of Eq.~(2), and averaging all the terms in the equation for the $L$ data samples, we can reach an identity of correlation matrices as

\begin{equation}
\hat{\mathbf{B}}=\hat{\mathbf{A}}\hat{\mathbf{C}}+\langle\Gamma(t)\mathbf{y}^{T}(t)\rangle,
\tag{S1}\label{eq:01}
\end{equation}
with
$${\hat{\mathbf{B}}=\langle{\dot{\mathbf{y}}(t)\mathbf{y}^{T}}(t)\rangle}\,,
\qquad{\hat{\mathbf{C}}=\langle{{\mathbf{y}}(t)\mathbf{y}^{T}}(t)\rangle},$$
\begin{equation}
{B_{ij}={\frac{1}{L}}{\sum_{q=1}^{L}}\dot{y}_{i}(t_{q})y_{j}(t_{q})}\,,
\qquad{C_{ij}={\frac{1}{L}}{\sum_{q=1}^{L}}y_{i}(t_{q})y_{j}(t_{q})},
\tag{S2}\label{eq:02}
\end{equation}
$${{\langle\Gamma(t)\mathbf{y}^{T}(t)\rangle\approx{\frac{1}{L}}{\sum_{q=1}^{L}}\Gamma(t'_{q})\mathbf{y}^{T}(t_{q})}}\,,\qquad{t_{q}<t'_{q}<t_{q}+\Delta{t_{q}}}.$$
Since noise $\Gamma(t'_{q})$ contributing to the velocity $\mathbf{\dot{y}}(t_{q})$ in Eq.~(4) (and also contributing to matrix $\hat{\mathbf{B}}$) is taken during $(t_{q},t_{q}+{\Delta}t_{q})$ and ${\Delta}t_{q}$ is much longer than the noise correlation time, $\Gamma(t'_{q})$ is thus not correlated with $\mathbf{y}^{T}(t_{q})$ and the second term in Eq.~(\ref{eq:01}) is vanishing. We thus reach a simple algorithm $\hat{\mathbf{B}}=\hat{\mathbf{A}}\hat{\mathbf{C}}$ leading to Eq.~(5), i.e.,

\begin{equation}
\hat{\mathbf{A}}=\hat{\mathbf{B}}\hat{\mathbf{C}}^{-1}.
\tag{S3}\label{eq:03}
\end{equation}

We can also multiply the both sides of Eq.~(2) by $\mathbf{y}^{+T}(t)=\mathbf{y}^{T}(t+{\Delta}t)$, and compute the corresponding correlations, and arrive at
\begin{equation}
\hat{\mathbf{B}}^{+}=\hat{\mathbf{A}}\hat{\mathbf{C}}+\langle\Gamma(t)\mathbf{y}^{+T}(t)\rangle,
\tag{S4}\label{eq:04}
\end{equation}

$${\hat{\mathbf{B}}^{+}=\langle{\dot{\mathbf{y}}(t)\mathbf{y}^{+T}}(t)\rangle}\,,
\qquad{B^{+}_{ij}={\frac{1}{L}}{\sum_{q=1}^{L}}\dot{y}_{i}(t_{q})y^{+}_{j}(t_{q})}={\frac{1}{L}}{\sum_{q=1}^{L}}\dot{y}_{i}(t_{q})y_{j}(t_{q}+{\Delta}t_{q})$$

$${\langle\Gamma(t)\mathbf{y}^{+T}(t)\rangle\approx{\frac{1}{L}}{\sum_{q=1}^{L}}\Gamma(t'_{q})\mathbf{y}^{+T}(t_{q})={\frac{1}{L}}{\sum_{q=1}^{L}}\Gamma(t'_{q})\mathbf{y}^{T}(t_{q}+{\Delta}t_{q})},$$
where $\mathbf{y}^{+}(t_{q})=\mathbf{y}(t_{q}+{\Delta}t_{q})$ is taken at time instant $t_{q}+{\Delta}t_{q}$, and thus is definitely correlated with $\Gamma(t'_{q})$ which is taken during $(t_{q},t_{q}+{\Delta}t_{q})$. This correlation can be exactly calculated in the continuous limit by solving Eq.~(2) as

$$\mathbf{y}(t)=e^{\hat{\mathbf{A}}t}{\int^{t}_{t_{0}}}{e^{-\hat{\mathbf{A}}t'}}{\Gamma(t')}dt',$$

$$\langle\Gamma(t)\mathbf{y}^{T}(t)\rangle=e^{\hat{\mathbf{A}}t}{\int^{t-0}_{t_{0}}}{e^{-\hat{\mathbf{A}}t'}}{\langle\Gamma(t)\Gamma^{T}(t')\rangle}dt'
=e^{\hat{\mathbf{A}}t}{\int^{t-0}_{t_{0}}}{e^{-\hat{\mathbf{A}}t'}}{\hat{\mathbf{Q}}{\delta(t-t')}}dt',$$

\begin{equation}
\langle\Gamma(t)\mathbf{y}^{+T}(t)\rangle=e^{\hat{\mathbf{A}}t}{\int^{t+0}_{t_{0}}}{e^{-\hat{\mathbf{A}}t'}}{\langle\Gamma(t)\Gamma^{T}(t')\rangle}dt'
=e^{\hat{\mathbf{A}}t}{\int^{t+0}_{t_{0}}}{e^{-\hat{\mathbf{A}}t'}}{\hat{\mathbf{Q}}{\delta(t-t')}}dt'.
\tag{S5}\label{eq:05}
\end{equation}
Due to the delta function of noise correlation in Eq.~(\ref{eq:05}), the integral is singular at time $t$, with $\mathbf{y}^{+}(t)$ ($\mathbf{y}(t)$) including (not including) the noise impact at time $t$. And thus we have
\begin{equation}
\langle\Gamma(t)\mathbf{y}^{T}(t)\rangle=0,
\tag{S6a}\label{subeq:1}
\end{equation}
\begin{equation}
\langle\Gamma(t)\mathbf{y}^{+T}(t)\rangle=\hat{\mathbf{Q}}.
\tag{S6b}\label{subeq:2}
\end{equation}
Substituting Eq.~(\ref{subeq:1}) to Eq.~(\ref{eq:01}) we obtain formula (5) or (\ref{eq:03}), and substituting Eq.~(\ref{subeq:2}) to (\ref{eq:04}) we have $\hat{\mathbf{B}}^{+}=\hat{\mathbf{A}}\hat{\mathbf{C}}+\hat{\mathbf{Q}}$ leading to,
\begin{equation}
\hat{\mathbf{A}}=(\hat{\mathbf{B}}^{+}-\hat{\mathbf{Q}})\hat{\mathbf{C}}^{-1},
\tag{S7}\label{eq:07}
\end{equation}
and also equivalently to
\begin{equation}
\hat{\mathbf{Q}}=\hat{\mathbf{B}}^{+}-\hat{\mathbf{B}}.
\tag{S8}\label{eq:08}
\end{equation}
Due to the identity
\begin{eqnarray*}
B^{+}_{ij}={\frac{1}{L}}{\sum_{q=1}^{L}}\dot{y}_{i}(t_{q})y^{+}_{j}(t_{q})={\frac{1}{L}}{\sum_{q=1}^{L}}{\frac{y^{+}_{i}(t_{q})-y_{i}(t_{q})}{\Delta{t_{q}}}}y^{+}_{j}(t_{q})
={\frac{1}{L}}{\sum_{q=1}^{L}}{\frac{y^{+}_{i}(t_{q})y^{+}_{j}(t_{q})-y_{i}(t_{q})y^{+}_{j}(t_{q})}{\Delta{t_{q}}}}
\\\approx{\frac{1}{L}}{\sum_{q=1}^{L}}{\frac{y_{i}(t_{q})y_{j}(t_{q})-y_{i}(t_{q})y^{+}_{j}(t_{q})}{\Delta{t_{q}}}}
={\frac{1}{L}}{\sum_{q=1}^{L}}{\frac{y_{j}(t_{q})-y^{+}_{j}(t_{q})}{\Delta{t_{q}}}}y_{i}(t_{q})=-{\frac{1}{L}}{\sum_{q=1}^{L}}\dot{y}_{j}(t_{q})y_{i}(t_{q})=-B_{ji}
\end{eqnarray*}
leading to Eq.~(6), i.e.,
$$\hat{\mathbf{B}}^{+}=-\hat{\mathbf{B}}^{T},$$
and
\begin{equation}
\hat{\mathbf{Q}}=-(\hat{\mathbf{B}}+\hat{\mathbf{B}}^{T})=-2\hat{\mathbf{B}}_{s}.
\tag{S9}\label{eq:14}
\end{equation}
With Eqs.~(5) and (6), we can depict both network interaction Jacobian $\hat{\mathbf{A}}$ and noise statistics $\hat{\mathbf{Q}}$ from the output data $\mathbf{y}(t)$ only, and reach the target of Fig.~1(d).

Actually, Eq.~(\ref{eq:14}) can be derived with another way. Substituting Eq.~(\ref{eq:03}) to the identity $\hat{\mathbf{A}}\hat{\mathbf{C}}+\hat{\mathbf{C}}\hat{\mathbf{A}}^{T}=-\hat{\mathbf{Q}}$ we can also obtain
$$\hat{\mathbf{Q}}=-(\hat{\mathbf{B}}+\hat{\mathbf{B}}^{T})=-2\hat{\mathbf{B}}_{s}.$$

\section{Inference of a periodic network}

\begin{figure}[h]
\renewcommand\thefigure{S\arabic{figure}}
\resizebox{12.0cm}{12.5cm}{\includegraphics{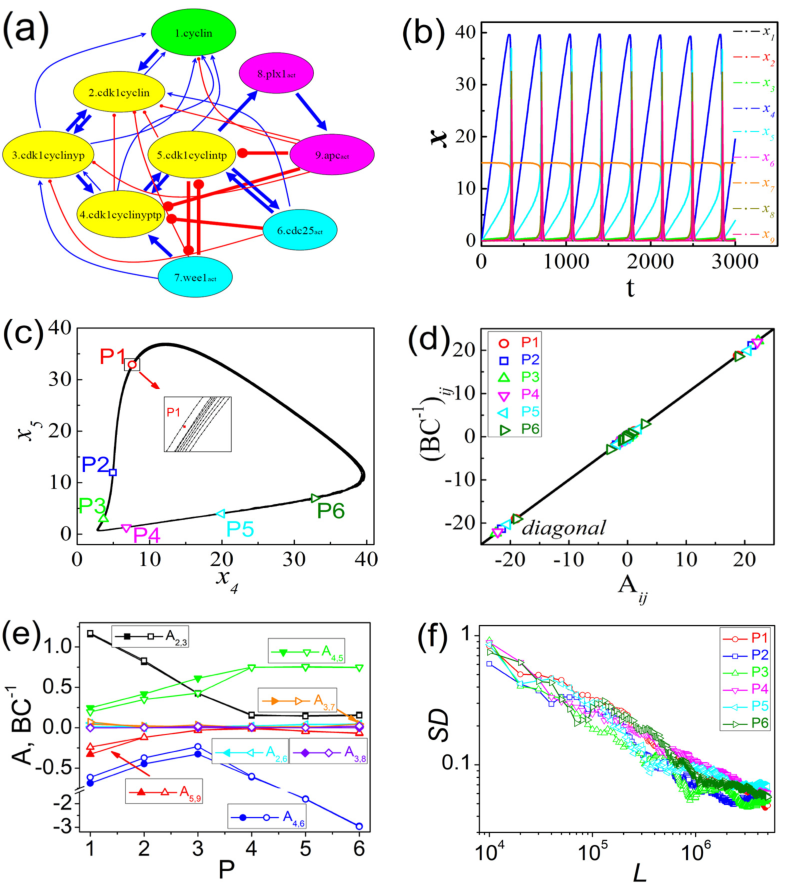}}
\caption{\label{Fig1} (Color online) Inverse computation of the DCM method on a cell cycle model \cite{Tsai} driven by white noise (Eqs.~(\ref{eq:09})). (a) The cell cycle regulatory network. Thick (thin) lines represent strong (weak) interactions. $\rightarrow$ represents active interactions and $\multimap$ repressive ones. (b) All the $9$ variable sequences produced by Eqs.~(\ref{eq:09}) available for the inverse computation. The motions of $9$ proteins are strongly heterogeneous, some of which have large variation amplitudes while some others show very weak oscillations. (c) A $2D$ projection of the trajectory of Eqs.~(\ref{eq:09}). (d) $[BC^{-1}]_{ij}$ plotted against actual $A_{ij}$ at six different phase points (shown in (c)) by applying the DCM method of Eq.~(5) to the measurable output data of (b)(c). $L=5\times{10^{5}}$ samples are yielded for inferences for each phase space region $P_{i}$. ($A_{ij}$)s are the actual Jacobian matrix elements at the corresponding points. All dots are located around the diagonal line, confirming the validity of the DCM method. (e) Some matrix elements computed by algorithm (5) (hollow dots) and the corresponding actual Jacobian values (solid dots) for the cell cycle system at six different phase space points. All the computed values coincide well with actual ones with certain small fluctuations. (f) The same as Fig.~2(f) with the cell cycle system considered. All samples are taken around the six space points indicated in (c), respectively. The standard deviations ($SD$s) of the DCM method decrease monotonously as $\frac{1}{\sqrt{L}}$ for all the six phase space points. }
\end{figure}

In Fig.~2 we studied systems linearized around a stable fixed point. Many dynamical networks may have time varying noise-free states (such as limit cycle, chaos and so on), then the full algorithms from Eq.~(4) to Eq.~(7) should be taken into account around various localized space domains. Here, we examine the DCM method through computational studies by depicting a model of CDK1 oscillations in the Xenopus embryonic cell cycle \cite{Tsai}. The model includes a negative feedback loop (active CDK1 brings about its inactivation through the anaphase promoting complex (APC)) and a pair of positive feedback loops (active CDK1 activates its activator Cdc25 and inactivates its inhibitor Wee1) (see Fig.~\ref{Fig1}(a)). The system is specified as a noise-driven dynamic network of nine variables (Eqs.~(\ref{eq:09})) with $\mathbf{x}(t)=[x_{1},x_{2},\ldots,x_{9}]=[cyclin(t), cdk1cyclin(t), cdk1cyclinyp(t), cdk1cyclinyptp(t), cdk1cyclintp(t), cdc25_{act}(t),\\ wee1_{act}(t), plx1_{act}(t), apc_{act}(t)]$,

\begin{equation}
\left\{
\begin{array}{l}
\frac{dx_{1}}{dt}=k_{synth}-k_{dest}x_{9}x_{1}-k_{a}(cdk1_{tot}-x_{2}-x_{3}-x_{4}-x_{5})x_{1}+k_{d}x_{2}+\Gamma_{1},\\
\frac{dx_{2}}{dt}=k_{a}(cdk1_{tot}-x_{2}-x_{3}-x_{4}-x_{5})x_{1}-k_{d}x_{2}-k_{dest}x_{9}x_{2}-k_{wee1}x_{7}x_{2}\\
~~~~~~~-k_{wee1basal}(wee1_{tot}-x_{7})x_{2}+k_{cdc25}x_{6}x_{3}+k_{cdc25basal}(cdc25_{tot}-x_{6})x_{3}+\Gamma_{2},\\
\frac{dx_{3}}{dt}=k_{wee1}x_{7}x_{2}+k_{wee1basal}(wee1_{tot}-x_{7})x_{2}-k_{cdc25}x_{6}x_{3}\\
~~~~~~~-k_{cdc25basal}(cdc25_{tot}-x_{6})x_{3}-k_{cak}x_{3}+k_{pp2c}x_{4}-k_{dest}x_{9}x_{3}+\Gamma_{3},\\
\frac{dx_{4}}{dt}=k_{cak}x_{3}-k_{pp2c}x_{4}-k_{cdc25}x_{6}x_{4}-k_{cdc25basal}(cdc25_{tot}-x_{6})x_{4}+k_{wee1}x_{7}x_{5}\\
~~~~~~~+k_{wee1basal}(wee1_{tot}-x_{7})x_{5}-k_{dest}x_{9}x_{4}+\Gamma_{4},\\
\frac{dx_{5}}{dt}=k_{cdc25}x_{6}x_{4}+k_{cdc25basal}(cdc25_{tot}-x_{6})x_{4}-k_{wee1}x_{7}x_{5}\\
~~~~~~~-k_{wee1basal}(wee1_{tot}-x_{7})x_{5}-k_{dest}x_{9}x_{5}+\Gamma_{5},\\
\frac{dx_{6}}{dt}=k_{cdc25on}(\frac{x_{5}^{ncdc25}}{{ec50_{cdc25}^{ncdc25}}+x_{5}^{ncdc25}})(cdc25_{tot}-x_{6})-k_{cdc25off}x_{6}+\Gamma_{6},\\
\frac{dx_{7}}{dt}=-k_{wee1off}(\frac{x_{5}^{nwee1}}{{ec50_{wee1}^{nwee1}}+x_{5}^{nwee1}})x_{7}+k_{wee1on}(wee1_{tot}-x_{7})+\Gamma_{7},\\
\frac{dx_{8}}{dt}=k_{plx1on}(\frac{x_{5}^{nplx1}}{{ec50_{plx1}^{nplx1}}+x_{5}^{nplx1}})(plx1_{tot}-x_{8})-k_{plx1off}x_{8}+\Gamma_{8},\\
\frac{dx_{9}}{dt}=k_{apcon}(\frac{x_{8}^{napc}}{{ec50_{plx1}^{napc}}+x_{8}^{napc}})(apc_{tot}-x_{9})-k_{apcoff}x_{9}+\Gamma_{9},\\
{\langle{\Gamma(t)_{i}}\rangle=0}, {\langle{\Gamma_{i}(t)\Gamma_{j}(t')}\rangle=Q_{ij}\delta(t-t')}.
\end{array}
\right.
\tag{S10}\label{eq:09}
\end{equation}
At parameters [$r=10, k_{synth}=0.16, k_{dest}=0.01, k_{a}=0.1, k_{d}=0.001, k_{wee1}=0.05, k_{wee1basal}=k_{wee1}/r, k_{cdc25}=0.1, k_{cdc25basal}=k_{cdc25}/r,cdk1_{tot}=230, cdc25_{tot}=15, wee1_{tot}=15, apc_{tot}=50, plx1_{tot}=50, nwee1=4, ncdc25=4, napc=4, nplx1=4, ec50_{plx1}=40, ec50_{wee1}=40, ec50_{cdc25}=40, ec50_{apc}=40, k_{cdc25on}=1.75, k_{cdc25off}=0.2, k_{apcon}=1, k_{apcoff}=0.15, k_{plx1on}=1, k_{plx1off}=0.15, k_{wee1on}=0.2, k_{wee1off}=1.75, k_{cak}=0.8, k_{pp2c}=0.008$] and initial condition [$\mathbf{x}(0)=(0,0,0,0,0,0,wee1_{tot},0,0)$] which are used in \cite{Tsai}, and with noise correlation $\hat{\mathbf{Q}}$ given as $\hat{\mathbf{Q}}_{ij}=\sigma_{i}\delta_{ij}, \sigma_{i}=0.0001$, we obtain the data set of variable sequences given in Fig.~\ref{Fig1}(b), of which the trajectory in a $2D$ phase plane is presented in Fig.~\ref{Fig1}(c). The data seemly show nice periodic orbit. The amplification in the small frame of Fig.~\ref{Fig1}(c) presents points in a small region which show clear disorder, and the dots scattered by noise perturbations provide necessary information to depict the interaction structure of Fig.~\ref{Fig1}(a).

Unlike the cases of Fig.~2, now the strengthes of interactions vary with phase space points, and the Jacobian matrix $\hat{\mathbf{A}}({\mathbf{x}})$ becomes $\mathbf{x}$ dependent. The tasks of inference are: ({\romannumeral1}) to determine the natures of $\hat{\mathbf{A}}$ elements (positive active regulations, negative repressive regulation or zero null links) which are $\mathbf{x}$-independent; and ({\romannumeral2}) to compute the detailed values of intensities of active and repressive interactions, which are definitely $\mathbf{x}$-dependent, from the data of Figs.~\ref{Fig1}(b) and \ref{Fig1}(c).

Among all the $81$ elements of $\hat{\mathbf{A}}$ there are $42$ null interactions, $25$ strong elements (thick arrowed lines in (a)) with absolute values larger than $0.30$, and other $14$ weak elements (thin arrowed lines) smaller than $0.10$. In Fig.~\ref{Fig1}(d) we arbitrarily choose six space points ($P_{1}-P_{6}$ shown in Fig.~\ref{Fig1}(c)) to infer interaction Jacobian matrices $\hat{\mathbf{A}}(\mathbf{x})$ by using algorithm (5). All the data locate closely near the diagonal line, indicating satisfactory inferences for all the six space points. Moveover, all the strong interactions are clearly distinguished from null interactions, with both their computed signs and strengths, satisfactorily coinciding with their actual signs and values. In Fig.~\ref{Fig1}(e) we plot several matrix elements computed at different phase space regions. The matrix elements vary with different reference places and fluctuate slightly due to noise. The values for null and very weak links always fluctuate around zero ($A_{3,8},A_{2,6},A_{3,7}$). Some strong matrix elements show values far away from zero at all phase space regions ($A_{2,3},A_{4,6}$), clearly showing their active and repressive natures. There are also some weak interactions with values not apparently deviating from null-links at certain points, however, they can clearly show their regulation natures, distinguishing from the null-links, in some other phase points ($A_{5,9},A_{4,5}$). Fig.~\ref{Fig1}(f) shows the standard deviation ($SD$) of Eq.~(8) against the sample numbers $L$ feasible for all the six phase space points. $SD$s of the DCM method scale as $\sqrt{\frac{1}{L}}$, similar to Fig.~2(f).

\section{Inference computations for networks with different dynamic state and different topology}

\begin{figure}[h]
\renewcommand\thefigure{S\arabic{figure}}
\resizebox{12.0cm}{4.5cm}{\includegraphics{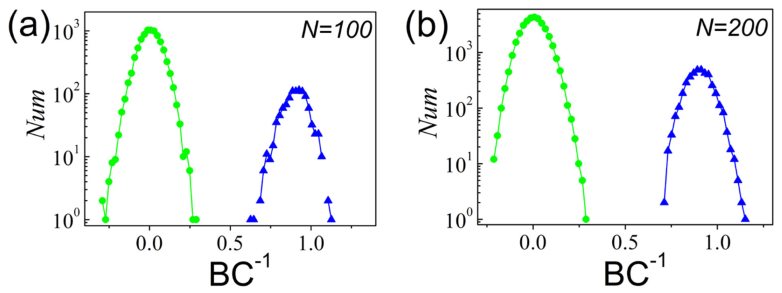}}
\caption{\label{Fig2} (Color online) Inverse computation for the chaotic R\"{o}ssler systems Eq.~(\ref{eq:10}) ($L=5\times{10^{5}}$). (a)(b) Randomly constructed asymmetric network structures (Links $A_{ij}(a)=1$ are randomly chosen with probability $p=0.10$ and $A_{ij}(0)=0$ for all other off-diagonal matrix elements). The noise statistics in Eq.~(\ref{eq:10}) is specified as $\hat{\mathbf{Q}}_{ij}=\sigma_{i}\delta_{ij}, \sigma_{i}=1.0$. Without noise the dynamic networks process synchronous chaos. The network structures can be successfully retrieved for different network sizes ((a) $N=100$ and (b) $N=200$).  }
\end{figure}

\begin{figure}[h]
\renewcommand\thefigure{S\arabic{figure}}
\resizebox{8.0cm}{6cm}{\includegraphics{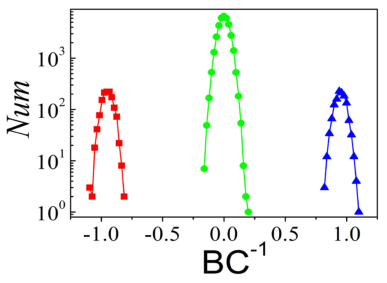}}
\caption{\label{Fig3} (Color online) Inverse computation for scale-free networks with asymmetric structure with approximate power law distributions for both in- and out-degree ($L=5\times{10^{5}}$). $N=200$, $\langle{k}\rangle=13$, and all the $A_{ij}$ construction and noise statistics are the same as Fig.~2(a) (except $A_{ii}=-5$). The DCM method can successfully depict the interaction Jacobian matrix $\hat{\mathbf{A}}$ with scale-free structure, and the figure is very similar to Fig.~2(b) with randomly distributed interactions. }
\end{figure}

A wide range of practical systems, including many biological and social systems, are modeled by dynamic complex networks described by Eq.~(1), and thus the DCM method can be generally used to solve inverse problems of realistic systems in various fields. The method can be applied to complex networks with different noise-free states, such as stable steady (Fig.~2), periodic (Fig.~\ref{Fig1}) and even chaotic states (Fig.~\ref{Fig2}); states with nonsynchronous (Fig.~\ref{Fig1}) and synchronous (Fig.~\ref{Fig2}) relationships between node motions; and to different topologies, such as random (Fig.~2) and scale-free (Fig.~\ref{Fig3}) networks.

In the main context and SM~\uppercase\expandafter{\romannumeral2} we studied the inverse problems of steady and nonsynchronous periodic networks driven by noise. The DCM method can be also applied to both synchronous and nonsynchronous chaotic noise-free states.

In Fig.~\ref{Fig2} we consider networks with synchronous chaotic noise-free state. We use the chaotic R\"{o}ssler system as the local node dynamics

\begin{equation}
\left\{
\begin{array}{l}
\dot{x}_{i}=-{y}_{i}-{z}_{i}+{\sum_{j=1}^{N}}A_{ij}({x}_{j}-{x}_{i})+\Gamma_{i}, \\
\dot{y}_{i}={x}_{i}+0.2{y}_{i}+{\sum_{j=1}^{N}}A_{ij}({y}_{j}-{y}_{i}), \\
\dot{z}_{i}=0.2+{z}_{i}\times({x}_{i}-9.0)+{\sum_{j=1}^{N}}A_{ij}({z}_{j}-{z}_{i}).
\end{array}
\right.
\tag{S11}\label{eq:10}
\end{equation}
The interaction structure $\hat{\mathbf{A}}$ is asymmetric and randomly constructed, as indicated in the caption of Fig.~\ref{Fig2}, and noise statistics are specified as $\hat{\mathbf{Q}}_{ij}=\sigma_{i}\delta_{ij}, \sigma_{i}=1.0$. It is again confirmed that the DCM method can satisfactorily infer all Jacobian matrix elements in this noise-driven chaotic network with synchronous noise-free state.

In Figs.~2,~\ref{Fig2} we studied networks with random interactions and found that the DCM method works well. The DCM method can work equally well for networks with different topologies. In Fig.~\ref{Fig3} we show that the DCM method can also effectively infer scale-free networks.

\section{Comparisons of DCM method with other three commonly-used inference methods}

The currently prevailing inference methods are based on different information included in the output data to infer network structures. These methods can be classified into several groups. Here, we introduce three commonly-used methods for comparisons.

(\romannumeral1) Pearson correlation coefficient $r$ is one of most commonly-used methods in biology for treating experimental data, due to its easiness of intuitive understanding and convenience of practical computations. In gene regulatory networks, matrix elements are calculated between all possible transcription factors $X$ and their possible target genes $Y$ as follows~\cite{Eisen1998}:
\begin{equation}
r(X,Y)=\frac{\sum_{i=1}^{n}(X_{i}-\langle{X}\rangle)(Y_{i}-\langle{Y}\rangle)}{\sqrt{\sum_{i=1}^{n}(X_{i}-\langle{X}\rangle)^{2}}
\sqrt{\sum_{i=1}^{n}(Y_{i}-\langle{Y}\rangle)^{2}}},
\tag{S12}\label{eq:11}
\end{equation}
where $n$ is the number of measurements.

(\romannumeral2) For two stochastic variables $X$ and $Y$, mutual information is defined as~\cite{Basso2005,Faith2007}:
\begin{equation}
MI(X,Y)=\sum_{i}\sum_{j}p(x_{i},y_{i})log\frac{p(x_{i},y_{i})}{p(x_{i})p(y_{i})}.
\tag{S13}\label{eq:12}
\end{equation}
In this implementation, variables $X$ and $Y$ represent an interacting node variable and a possible target node variable, respectively.

(\romannumeral3) The regression method is one of the most commonly-used method in solving the inverse problems~\cite{Haury}. For steady-state data of Eq.~(2), the variable of the target node $y_{i}$ is modeled as a linear function of the variables of all possible interacting nodes as
\begin{equation}
y_{i}=\sum_{j=1,j\neq{i}}^{N}\beta_{ij}y_{j}+\xi_{i}.
\tag{S14}\label{eq:13}
\end{equation}
And the target of minimal difference error between Eq.~(\ref{eq:13}) for assumed matrix $\hat{\beta}$ and the the measurable data is pursued in the regression computation tests for depicting Jacobian matrix elements.

\begin{figure}[h]
\renewcommand\thefigure{S\arabic{figure}}
\resizebox{12.0cm}{9.0cm}{\includegraphics{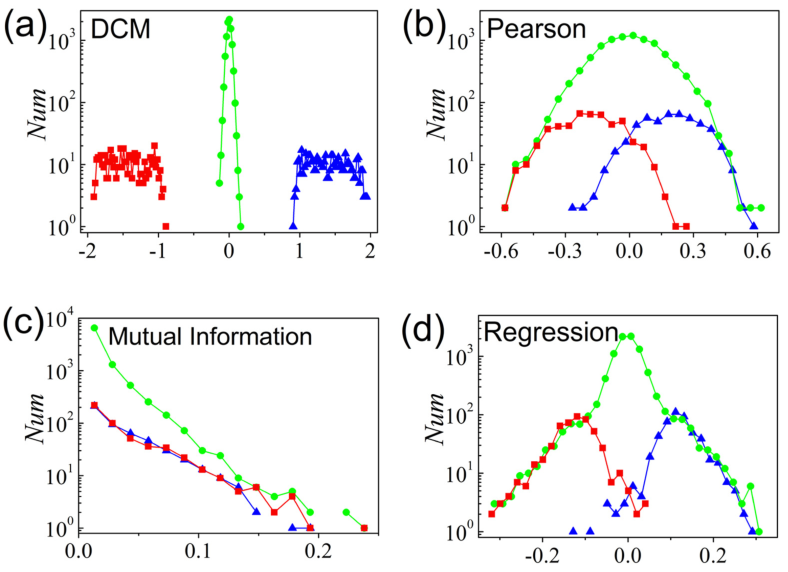}}
\caption{\label{Fig4} (Color online) Comparison of the DCM method with three commonly-used inference methods introduced in SM~\uppercase\expandafter{\romannumeral4}. The same data set of Case 2 of Fig.~2 are used for all the four methods: DCM method (a); Pearson correlation (b); Mutual information (c) and Regression (d) (all colors have the same meaning as Figs.~2(b)(d)). It is demonstrated that the DCM  method is a unique method to depict network structures, from output data only, for both interaction natures and coupling strengths. }
\end{figure}

\begin{figure}[h]
\renewcommand\thefigure{S\arabic{figure}}
\resizebox{12.0cm}{9cm}{\includegraphics{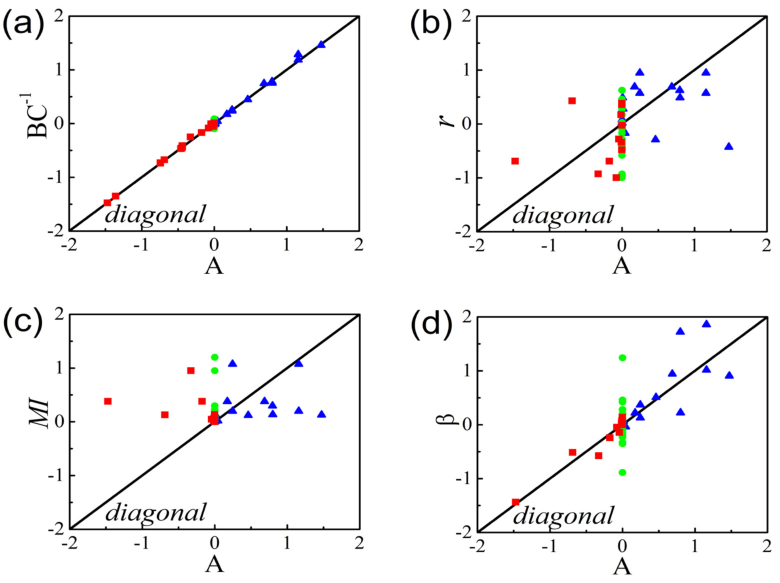}}
\caption{\label{Fig5} (Color online) The same as Fig.~\ref{Fig4} with the cell cycle model Eq.~(\ref{eq:09}) considered. The data set of $P_{1}$ in Fig.~\ref{Fig1}(c) are used for all the four methods: DCM method (a); Pearson correlation (b); Mutual Information (c) and Regression (d). In (a)-(d) the DCM  method is a unique method to depict network structures, from output data only, for both interaction natures and coupling strengths. The dots of all the three commonly-used methods deviate from the diagonal line considerably, indicating incorrect inferences of interaction strengths. In particular, some wrong conclusions of interaction natures can be observed in (b)(c)(d).}
\end{figure}

In Fig.~\ref{Fig4} we compare the DCM method with all the above three methods by using the data set of Case 2 in Fig.~1, and in Fig.~\ref{Fig5} we do the same as Fig.~\ref{Fig4} with the model Eq.~(\ref{eq:09}) considered (the inverse computations are conducted around the phase space point $P_{1}$). In both figures, the DCM method demonstrates obvious advantages in correctly depicting natures and strengths of interactions over all the other three commonly-used methods.

\bibliography{apssamp2}

